*Subject Section*

# ACCORDION: Clustering and Selecting Relevant Data for Guided Network Extension and Query Answering

Yasmine Ahmed[1, *], Cheryl Telmer[2] and Natasa Miskov-Zivanov[1,3, *]

[1]Electrical and Computer Engineering Department, [3]Bioengineering, Computational and Systems Biology, University of Pittsburgh, [2]Department of Biological Sciences, Carnegie Mellon University, Pittsburgh, PA

*To whom correspondence should be addressed.

## Abstract

Querying new information from knowledge sources, in general, and published literature, in particular, aims to provide precise and quick answers to questions raised about a system under study. In this paper, we present ACCORDION (**A**utomated **C**lustering **C**onditional **O**n **R**elating **D**ata of **I**nteractions t**O** a **N**etwork), a novel tool and a methodology to enable efficient answering of biological questions by automatically assembling new, or expanding existing models using published literature. Our approach integrates information extraction and clustering with simulation and formal analysis to allow for an automated iterative process that includes assembling, testing and selecting the most relevant models, given a set of desired system properties. We applied our methodology to a model of the circuitry that controls T cell differentiation. To evaluate our approach, we compare the model that we obtained, using our automated model extension approach, with the previously published manually extended T cell differentiation model. Besides demonstrating automated and rapid reconstruction of a model that was previously built manually, ACCORDION can assemble multiple models that satisfy desired properties. As such, it replaces large number of tedious or even impractical manual experiments and guides alternative hypotheses and interventions in biological systems.

**Contact:** {yaa38, nmzivanov}@pitt.edu
**Supplementary information:** Supplementary data are available at *Bioinformatics* online.

## 1 Introduction

While modeling helps explain complex systems, guides data collection and generates new challenges and questions [1], it is still largely dependent on human contributions. For example, in biology, model creation requires reading hundreds of papers, extracting useful information manually, incorporating background and common-sense knowledge of domain experts, and conducting wet-lab experiments. These time-consuming steps make the creation and the development of models a slow, laborious and error-prone process. In addition, as the amount of biological data in the public domain grows rapidly, problems of data inconsistency and fragmentation are arising [2]. Therefore, the automation of model building, and even more, of model extension, when new information becomes available, or when the domain knowledge advances, is a critical next step for computational modeling. Such automation will not only lead to more efficient modeling due to reducing the amount of slow human interventions, but will also allow for more consistent, comprehensive and robust modeling process.

In the last few decades, computer models have been used to explain how biomolecular signaling pathways regulate cell functions. Usually, modelers start with a few seed components and their interactions, or with a baseline model that can be found in curated public model databases such as Reactome [3], STRING [4], KEGG [5], or in published literature. Depending on the questions to be answered with modeling, the baseline model is usually further extended with the information extracted from published literature or obtained from domain experts [6]. In order to automate the collection of articles and information extraction, one begins with a formal search query, which is defined according to a question posed about the modeled system. The search query guides automated selection of articles that contain relevant information from published literature databases. As the biomedical literature mining tools are becoming essential for the high throughput extraction of knowledge from scientific papers, we use in our work existing machine reading engines. We then use the extracted information to extend or assemble models in order to answer questions about the system under investigation [7].

In [8], the authors proposed a method that starts with a baseline model and selects interactions automatically extracted from published work. The goal of [8] was to build a model that satisfies pre-defined requirements or to identify new therapeutic targets, formally expressed as existing or desired system properties. As results in [8] demonstrate, automatic model extension is a promising approach for accelerating modeling, and consequently, disease treatment design. The authors in [8] organize the information extracted from literature into layers, based on their proximity to the baseline model. Recently, another extension method that uses a Genetic Algorithm (GA) was proposed in [9]. The GA-based approach was able to extract a set of extensions that led to the desired behavior of the final extended model. The disadvantages of the GA-based approach include non-determinism, as the solution may vary across multiple algorithm executions on the same inputs, as well as issues with scalability.

In this work, we propose ACCORDION (**A**utomated **C**lustering **C**onditional **O**n **R**elating **D**ata of **I**nteractions t**O** a **N**etwork), a tool that automatically and efficiently assembles the information extracted from available literature into models, tests the newly assembled models, and selects the most suitable model to address user questions. In contrast to [8], our approach focuses on identifying clusters of strongly connected elements in the newly extracted information, that have a measurable impact when added to the model. Once the interactions extracted from the literature are clustered, we score their performance on a selected set of system properties, using stochastic simulation methods [11] and statistical model checking [10]. The scoring helps determine which clusters to add to the baseline model. Therefore, ACCORDION takes at most a few hours to execute thousands of experiments *in silico*, which would take days, or months, or would be impractical to conduct *in vivo* or *in vitro*.

ACCORDION is versatile and can be used to extend many different models. We have selected a computational model of T cell differentiation [12] to demonstrate the accuracy, efficiency and utility of the tool. Our main goal with this case study is to show that ACCORDION is able to expand automatically, without human intervention, an existing published model into another published and manually built model, using new elements and new interactions automatically extracted from published literature. As the final golden model, we used the T cell model published in [13] and the set of desired system properties discussed in [12][13]. The golden model and the properties are used to evaluate the ACCORDION output. To this end, the contributions of this work include: (i) a new method to extend models by combining clustering and path finding with model testing on a set of formally written desired system properties; (ii) an evaluation of the effect of published literature and machine reading when automatically reproducing a manually built model; (iii) several new candidate models of the circuitry controlling naïve T cell differentiation, assembled automatically, satisfying the same set of desired properties as existing manually built models, and thus, enabling exploration of redundancies or discovering alternative pathways of regulation.

## 2 Background

We provide in this section an overview of several tools and background concepts that are used by ACCORDION. We first describe the tools that we have used to automatically find and read published papers relevant for user queries (Section 2.1). In order to use the extracted information in models, while retaining all the useful information, a suitable representation format for model components is critical (Section 2.2). We detail the components of the representation format and provide a brief overview of the model analysis techniques. ACCORDION uses these techniques and tools to evaluate the newly expanded models, and to select the best model to address user questions (Section 2.3).

### 2.1 Information extraction from literature

Extraction from literature usually starts with a question, for example, "How is PTEN regulation involved in T-cell fate?" We can write these questions as logical expressions (**Figure 1**(a)(left)). These formally written queries are used to search public literature databases (e.g., PubMed [15]) as illustrated in **Figure 1**(a)(middle). Once the relevant papers are selected, they are sent to machine reading engines for automated extraction of information (**Figure 1**(a)(middle)).

The state-of-the-art automated reading engines [18][19] are capable of finding hundreds of thousands of events in cellular signaling pathways from thousands of papers, in a few hours. Events in the machine reading output represent interactions between biochemical entities, such as post-translational modifications (e.g., binding, phosphorylation, ubiquitination, etc.), transcription, translation, translocation, and increase or decrease of amount or activity. In the context of biomedical literature, entities are usually proteins, chemicals, genes, and RNAs, although sometimes they also represent biological processes. For each extracted entity, reading engines provide its name, the database where it is characterized, and the database identifier (ID) for the entity. Machine reading also collects the evidence, usually a sentence from which the event was extracted. For our case study, we used an open-source reading engine, REACH [19], to quickly obtain information from biomedical literature. In **Figure 1**(a)(right), we show two example sentences. The REACH reading engine extracts events into an interaction-based format shown in **Figure 1**(b). We will refer to the list of interactions retrieved from literature in this format as *reading output*.

### 2.2 Model representation and executable models

The three rows in the table in **Figure 1**(b) can be automatically translated into the element-based BioRECIPES format [20], which is then used as input to the executable model generation (see Section 2.3). The BioRECIPES tabular model representation format is illustrated in **Figure 1**(c) with several examples of molecules and interactions in T cells [12]. In the examples, PTEN is positively regulated by Foxp3, and negatively regulated by TCR. Ras has one positive regulator, TCR, and no negative regulators. IL-2 has positive and negative regulators, Ras and Foxp3, respectively.

The BioRECIPES representation format includes, for each model element: (i) name, (ii) type (protein, gene, RNA, or a chemical), (iii) identifier from a database (e.g., UniProt [21]), (iv) variable that represents state, and (v) set of regulators. While the BioRECIPES format is a sufficient representation for all the relevant element and interaction information, all interactions in a model can also be represented as a directed graph $G(V, E)$, with a set of nodes $V$ and a set of directed edges $E$. Each node $v \in V$ corresponds to one model element, and each edge $e(v_i, v_j) \in E$ represents a directed interaction in which element $v_i$ regulates element $v_j$. The graphical representation of all model interactions is often referred to as an influence map, and it is especially useful for the methods that are used in ACCORDION, as will be discussed in Section 3.2. Next to the table in **Figure 1**(c), we show a graph of element interactions that are listed in the table. As can be seen in the graph, we include the information about the sign of the interaction in the form of arrow type, a pointed arrow represents positive regulation, while a blunt arrow represents negative regulation.

We will refer to the set of regulators of an element as its influence set, distinguishing between positive and negative regulators. Additionally, we can define a vector of all variables representing states of model elements as $x = (x_1, .., x_N)$, where $N=|V|$ is the total number of model elements. If we use Boolean variables, then $x_i \in \{0, 1\}$, where $i=1..N$. Next, we can assign a state transition function to any model element, which defines a state change of the element, given the states of its regulators. We will refer to these functions as *element update rules* and to the model with update rules as an *executable model*. In the case of Boolean variables representing element states, the basic operations are AND (*), OR (+) and NOT (!). For example, one version of update rules for the small graph in **Figure 1**(c) can be: PTEN = Foxp3 * !TCR, Ras = TCR, and IL-2 = Ras * !Foxp3. The choice between AND and OR operation depends on the available information about interactions and element regulations. For example, for an element to be "activated", all necessary regulators are combined with an AND operation, and all sufficient regulators with an OR operation.

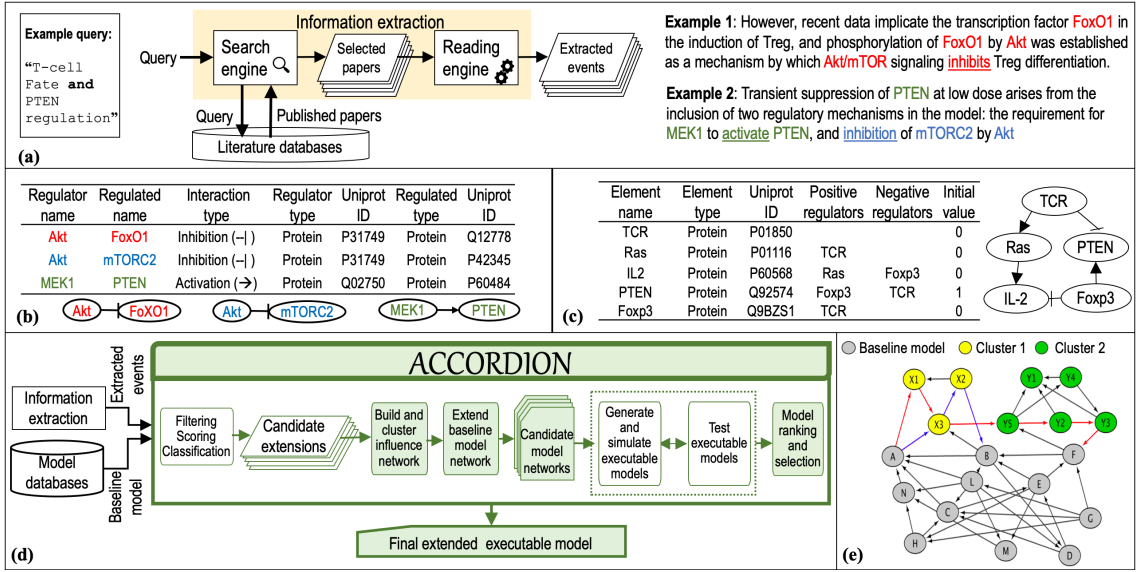

**Figure 1.** ACCORDION inputs and methodology overview: (a) **Left**: example query used to select relevant papers. **Middle**: main components of information extraction from relevant papers. **Right**: two example sentences with highlighted entities and events that are extracted by machine readers. (b) **Top**: tabular outputs from REACH engine when reading example sentences from (a). **Bottom**: graphical representation of REACH outputs. (c) **Left**: tabular representation of several elements and their influence sets (positive and negative regulators) in BioRECIPES format. **Right**: graphical representation of elements and influence sets. (d) The flow diagram of the ACCORDION processing steps, inputs, and outputs. (e) Toy example: a baseline model and connected clusters: blue edges highlight a return path within one cluster, and red edges show a return path connecting two clusters.

### 2.3 Model analysis
We describe here two methods that we use to analyze the models extended with the newly obtained information and data.

*2.3.1 Stochastic simulation.* We use the DiSH simulator [11] to observe dynamic behavior of the baseline model and the extended models. DiSH can simulate networks with multi-valued elements in both deterministic and stochastic manner, and we utilize both of these features in our analysis, as shown later in Section 5. Each simulation run starts with a specified initial model state, where initial values are assigned to all model elements to represent a particular system state (e.g., naïve T cell, regulatory T cell, etc.). Next, we use element update rules to determine element state transitions. We track element changes for a pre-defined number of simulation steps, or until a steady state is reached [11].

Here, we provide a brief description of the simulation approach, more details can be found in [11]. Furthermore, this approach has been used for simulation and analysis of discrete models, such as those described in [22][12], and was previously incorporated into several other tools [23][24][25]. If we assume that a simulation run has $M$ steps, we define a trajectory of element $x_i$ in the $k$th run as a time course of its state values $T_k(x_i) = (x_{i0}, x_{i1}, …, x_{iM})_k$ in time steps $t=0,…,M$. When the simulator is in the stochastic mode, in each simulation step, only one element is randomly chosen, and its new value is computed according to its update rule. Depending on the information available, the rates at which elements are updated can be different across model elements; when there is limited information about elements, we choose to use the same update rate for all elements. In either case, due to the randomness in element update order, multiple runs that start with the same initial state may result in different (non-deterministic) state transitions, and thus, in different trajectories of state changes in time. DiSH simulations output a file that includes all the simulated trajectories for all model elements, in other words, for $K$ runs, for each model element $x_i$, we obtain its simulated trajectories $T(x_i)=\{T_1(x_i), T_2(x_i)…, T_K(x_i)\}$. These trajectories can be used to plot and visualize behavior over time for any given element. Typically, averaged trajectories are plotted [12][22], where an average element state value is computed across all trajectories, in each simulation step.

*2.3.2 Statistical model checking.* In this work, we use statistical model checking [10][14] to test all generated models against formally defined properties. Model checking is often used to verify whether a model of a system, or a system design, satisfies a set of properties describing expected behavior of the system. Each property is encoded into Bounded Linear Temporal Logic (BLTL) [26][14]. Here, we use *statistical* model checking since the state transitions are not necessarily deterministic, and we follow the simulation approach described in Section 2.3.1. To avoid a full state space search, statistical model checking conducts randomized sampling to generate simulation trajectories of the model and performs statistical analysis on those trajectories. The input to the statistical model checker is a system property expressed as a BLTL formula, and the output is a *probability estimate* ($P$) that the model satisfies a given property, under particular error interval for the estimate. For instance, let us assume that we would like to test a property that, at any point within the first $s_1$ time steps, element $v_i$ becomes 1 and element $v_j$ becomes 0, and that they both keep those values for at least $s_2$ time steps. We would then write the formula: $F^{s_1} G^{s_2}(v_i = 1 \wedge v_j = 0)$, where $F^{s_1}$ stands for "any time in the future $s_1$ steps", and $G^{s_2}$ stands for "globally for $s_2$ steps".

## 3 Proposed methodology
The steps and components within ACCORDION are outlined in **Figure 1**(d). The first step of our proposed methodology is creating an input for

ACCORDION, which includes extracting new event information from literature by machine reading engines, followed by filtering, scoring and classifying these events. Once the new input is created, the three main steps within ACCORDION are performed, and they include (1) clustering of new events, (2) assembly of the clustered event data into models, and (3) selection of the most suitable and useful events. In the following subsections, we discuss each of these steps in detail.

### 3.1 Extraction and classification of new event information

To query for new information from literature databases, we write questions as search terms in the form of logical expressions, which can be used by literature search engines, either Non-Medline (e.g., Google), or Medline search tools (e.g., PubMed [15], Ovid [29]). The search engines return a list of papers most relevant to these search terms. The selected papers are then used as an input for machine reading engines, which extract entities and events from the papers. Once the event information is extracted, it is forwarded to the event classification tool, to identify potential extensions for the existing model. However, the output of machine reading engines can contain inconsistencies and errors. Therefore, extracted event information needs to be filtered before it can be used in models.

First, we select from the reading output the protein-protein interactions, not including the more general biological processes information. The rationale behind this is that there is often not enough context for a mentioned biological process, and the lack of context affects interpretation of the extracted interaction. The machine reading output is further filtered using public protein interaction databases [21][4][30], which increases the confidence in the interactions that will be used as potential extensions for models. To classify the remaining interactions, we use an interaction classification tool [31]. As described in Section 1, we assume that, in order to answer a query, we would most often start from an existing baseline model, and thus, the extracted interactions are classified according to their relationship with the baseline model. We classify interactions from the reading output into three groups: (a) *corroborations*, when the interaction matches an interaction that already exists in the model; (b) *contradictions*, when the interaction represents a contradicting regulatory mechanism from the one that exists between the same elements in the model; (c) *extensions*, when the interaction is not found in the model.

As corroborations confirm what is already in the baseline model, we do not use them in extending the baseline model. In our future work, we plan to include a confidence measure for the interactions in the model, and the corroborations found in literature would contribute to computing the confidence. Additionally, ACCORDION currently does not examine and utilize the information within the extracted contradictions, although they may hold useful information about the modeled system. In some cases, contradictions could be even considered as model extensions. For example, in the reading output, we often come across interactions stated as "A positively regulates B" or just "A regulates B", while the model includes interaction "A inhibits B" or "B inhibits A" or "B regulates A". Given that extracted contradictions can be further explored and the information within contradictions can sometimes lead to model improvements, we will explore them more carefully in our future work. To extend the baseline model, only the interactions that are classified as extensions form an input for ACCORDION, and in the rest of the paper, we will refer to these interactions as *Candidate Extension Interactions (CEIs)*.

### 3.2 Clustering of candidate extension interactions

The method used to identify clusters of extracted, filtered and classified CEIs is formally outlined in Algorithm 1 (see Figure S1 in the supplementary material) and described in detail here.

The set of CEIs can be represented as a set of candidate extension edges $E^{ext}$, and the source and target nodes of these edges that are not already in the baseline model graph, $G^{BM}(V^{BM}, E^{BM})$, will be members of the set of candidate extension nodes $V^{ext}$. We then create a new graph $G^{new}(V^{new}, E^{new})$, where $V^{new} = V^{BM} \cup V^{ext}$, and $E^{new} = E^{BM} \cup E^{ext}$. **Figure 1**(e) shows a toy example graph $G^{new}$, where grey nodes belong to the baseline model, while yellow and green nodes belong to the CEIs obtained from machine reading. We further classify the edges $e(v_s, v_t)$ from the set $E^{ext}$, where $v_s$ is the source node and $v_t$ is the target node, into the following categories: (i) edges in which both the source node $v_s$ and the target node $v_t$ belong to the baseline model: $\{v_s, v_t\} \in V^{BM}$; (ii) edges in which either a source node or a target node belongs to the baseline model: $(v_s \in V^{BM}$ and $v_t \notin V^{BM})$ or $(v_s \notin V^{BM}$ and $v_t \in V^{BM})$; (iii) edges in which neither the source node nor the target node belongs to the baseline model: $\{v_s, v_t\} \notin V^{BM}$.

Adding the CEIs to the baseline model all at once usually does not result in a useful and accurate model. Alternatively, we can add one interaction at a time and test each model version, which is time consuming, or even impractical, given that the number of models increases exponentially with the number of CEIs. Moreover, adding individual interactions does not have an effect on the model when an interaction belongs to category (iii), and most often when it belongs to category (ii). It proves much more useful to add paths of connected interactions, which are at the same time connected to the baseline model in at least two elements. Therefore, our approach for finding the most useful subset of the CEIs includes finding connected interactions, that is, a set of edges in the graph $G^{new}$ that form a return path. If we define a path of $p$ connected edges as $e^{path}(v_{s1}, v_{tp}) = (e_{i1}(v_{s1}, v_{t1}), e_{i2}(v_{s2}=v_{t1}, v_{t2}), e_{i3}(v_{s3}=v_{t2}, v_{t3}), \ldots, e_{ip}(v_{sp}=v_{tp-1}, v_{tp}))$, we will call $e^{path}(v_{s1}, v_{tp})$ a return path, when $\{v_{s1}, v_{tp}\} \in V^{BM}$ and $v_{s1} \neq v_{tp}$. In **Figure 1**(e), we highlight one such return path in blue. To find these return paths formed by CEIs, we conduct clustering of graph $G^{new}$ that includes both the baseline model and the CEIs.

To cluster CEIs, we use Markov Clustering algorithm (MCL) [32], an unsupervised graph clustering algorithm, commonly used in bioinformatics (e.g., clustering of protein-protein interaction networks [33][34]). In [35], the authors showed that the MCL algorithm is tolerant to noise, while identifying meaningful clusters. MCL is compared with, Affinity Propagation (AP) clustering algorithm, proposed in [36], and it is demonstrated that the MCL algorithm outperforms the AP algorithm [35]. Moreover, the analysis in [33] supported the superiority of MCL over other clustering techniques [37][38][39] in identifying protein complexes from interaction networks. MCL simulates random walks on an underlying interaction network, by alternating two operations, expansion and inflation. First, self-loops are added to the input graph representing biological interactions which is in our case graph $G^{new}$, and this graph is then translated into a stochastic Markov matrix [40]. This matrix consists of transition probabilities between all pairs of the graph nodes, and the probability of a random walk of length $p$ between any two nodes can be calculated by raising this matrix to the exponent $p$, a process called expansion. As longer paths are more common between nodes within the same cluster than between nodes across different clusters, the transition probabilities between nodes in the same cluster will typically be higher in these newly obtained expanded matrices. MCL further amplifies this effect by computing entry-wise exponents of the expanded matrix, a process called inflation [32], which raises each element of the matrix to the power $r$. Clusters are determined by alternating expansion and inflation, until the graph is partitioned into subsets such that there are no paths between these subsets.

### 3.3 Assembly of new interaction data into models

After generating clusters, the next step is to add them to the model. Similar to the discussion about individual extensions in Section 3.2, we can add clusters one at a time, or in groups. The more cluster or cluster groups we generate, the more models we need to assemble and test. Moreover, the number of possible cluster combinations grows with the total number of generated clusters, and the number of clusters depends on the inflation parameter *r*, as it directly influences cluster granularity [32]. To alleviate the problem of the large number of cluster combinations, we propose a method for combining the clusters found by the MCL algorithm. Formally, if the clusters we generated in the previous step are $C_1,…,C_n$, and we find a subset of clusters $C_{i1},..,C_{ij}$, where $1 < j \le n$, for which at least one return path exists that goes through all the clusters, then we merge these clusters into a single cluster that will be added to the model. An example of a multi-cluster path is highlighted in red in **Figure 1**(e), starting at *BM* (baseline model), connecting to $C_1$ (cluster 1), then connecting to $C_2$ (cluster 2), and from $C_1$ connecting back to *BM*. Therefore, we extend the baseline model with multiple clusters simultaneously, based on how clusters are connected to the model. The cluster merging procedure is outlined in Algorithm 2 (see Figure S2 in the supplementary material). Finally, we rank and score the final list of candidate clusters, based on the existence of return path, to choose the ones that will be used in model extension.

Next, we can select one or more clusters from the set of ranked and scored clusters to generate multiple *Candidate Executable Models (CEMs)*. Each CEM contains elements from both the baseline model and the selected cluster(s). Both procedures, the assembly of the CEI set, and the generation of CEMs are fully automated. However, element update rules are not necessarily unique, as previously discussed in [8]. For example, in the case of a Boolean model, if the original rule is "A = B + C", and the candidate extension states that "D positively regulates A", then the new update rule for A can be either "A= (B + C) * D", or "A= B + C + D". We will investigate the effect of adding a new regulatory element in both cases, using either AND (*) or OR (+) operation.

### 3.4 Selection of final extended model

We use both simulation and formal analysis to evaluate the CEMs. In order to simulate a model, all model elements need to be assigned a starting state (i.e., initial value). The initial values for the baseline model elements (nodes in the set $V^{BM}$) are typically already known, however, the newly added elements (nodes in the set $V^{ext}$) need to be assigned initial values as well. Unfortunately, machine reading does not usually provide this information. In this work, we assume that all elements within the same cluster have the same initial value. In Section 6, we will compare models with different initializations of the newly added elements to evaluate the effect of initialization on the behavior of the CEMs.

To obtain dynamic traces of the baseline model and the CEMs, we use the DiSH simulator (Section 2.3.1). We test each CEM using the statistical model checking approach (Section 2.3.2), by computing a probability estimate *P* for each property in a given set of system properties. As discussed in Section 2.3.2, the statistical model checker calls the simulator in order to obtain element trajectories for a defined number of steps. Assuming independence across system properties, for a given CEM, we compute the *global CEM probability* by multiplying the *P* values for all properties for the given CEM. Finally, we select the CEM that has the highest probability of satisfying the selected properties as our final extended executable model. The procedure for selecting this final CEM is summarized in Algorithm 3 (see Figure S3 in the supplementary material).

## 4 Case study: T cell differentiation

Naïve peripheral T cells are stimulated via antigen presentation to T cell receptor (TCR) and with co-stimulation at CD28 receptor. This stimulation results in the activation of several downstream pathways, feedback and feedforward loops between pathway elements, which then lead to the differentiation of naïve T cells into helper (Th) or regulatory (Treg) phenotypes. The distribution between Th and Treg cells within the T cell population depends on antigen dose; for instance, high antigen dose results in prevalence of Th cells, while low antigen dose leads to a mixed population of Th and Treg cells. The key markers that are commonly used to measure the outcomes of the naïve T cell differentiation into Th and Treg cells are IL2 and Foxp3, respectively. In other words, Th cells are characterized by high expression of IL-2 and low expression of Foxp3, and Treg cells are characterized by high expression of Foxp3 and low expression of IL-2. To demonstrate our model extension procedure, we use two existing, manually built models of T cell differentiation, from [12] and [13].

### 4.1 Baseline model and golden model

In [12], the authors proposed a model where most of the elements are assumed to have two main levels of activity, and are therefore represented with Boolean variables, and their update rules are logic functions. Additionally, the stimulation through TCR is assumed to have three different levels, no stimulation (TCR=0), low dose (TCR=1), and high dose (TCR=2), and therefore, it is implemented using two Boolean variables. We used the model from [12] to create the baseline model for our case study. The interaction map of this model is provided in [12] (also included in the supplementary material).

In [13], the authors have proposed an extension of the original T cell model from [12], a new model that improved the behavior of the original model. Specifically, in the new model in [13], the Foxp3 response to low dose is closer to experimental observations, that is, it is present in almost 70% of the differentiated population, while in [12] Foxp3 was present in 100% of the differentiated population. In both models, there is a brief transient induction of Foxp3 after the stimulation with high antigen dose. We will refer to the model from [13] as the golden model. For the baseline, we used the original model from [12], without several interactions overlapping with the golden model from [13] (TCR activates PIP3, PIP3 activates Akt, Akt activates mTORC2 and mTORC2 inhibits Akt). While the model from [12] satisfied a large number of system properties, except for a few that are satisfied by the model in [13] only, the baseline model in its reduced shape does not satisfy a larger set of system properties. Our aim is to use ACCORDION to automatically expand this baseline model in order to recapitulate the behavior of the golden model.

### 4.2 Set of properties

From the golden model in [13] and the results of its studies, we define a set of properties that our final automatically extended model needs to satisfy. Specifically, the properties capture observed responses of key pathway components in T cells, Foxp3, IL-2, PTEN, CD25, STAT5, AKT, mTOR, mTORC2 and FoxO1, to three scenarios: (1) no stimulation (TCR=0), (2) stimulation with low antigen dose (TCR=1), and (3) stimulations with high antigen dose (TCR=2). The complete list of 27 properties is shown in Figure S4 in the supplementary material.

## 5 Results

Here, we discuss several experiments that we conducted, and the results and our observations from these experiments.

### 5.1 Experimental setup

All models that we use or create are written in the BioRECIPES representation format, which was presented in [20] and briefly described in Section 2.2. From this format, executable models are generated automatically as part of the DiSH simulator [11], which is publicly available at [41].

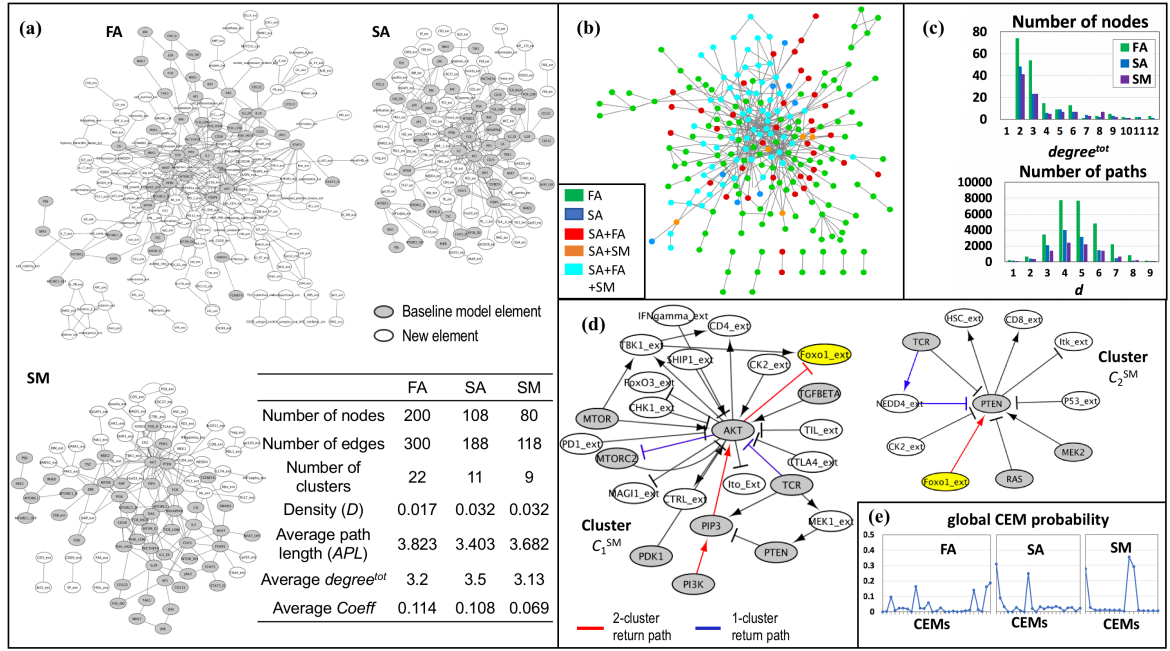

**Figure 2.** T cell model extension results. (a) Networks obtained when combining baseline model with the CEI set for each of the three cases, FA, SA, and SM ($G^{new,FA}$, $G^{new,SA}$, and $G^{new,SM}$). Gray nodes are the baseline model nodes and white nodes are the new nodes that belong to the CEIs. **Table**: Common graph features of $G^{new,FA}$, $G^{new,SA}$, and $G^{new,SM}$. (b) $G^{new,FA}$, $G^{new,SA}$, and $G^{new,SM}$ drawn together, highlighting the common nodes. (c) **Top**: histogram of $degree^{tot}$ values and corresponding number of nodes; **Bottom**: histogram of node distance ($d$) values and the corresponding number of paths. (d) Two clusters that form a return path with the baseline T cell model, shown as directed graphs (yellow node is a common node for both clusters). (e) Global (overall) probability of satisfying desired system properties for each candidate extension model (CEM) that ACCORDION assembled from each of the three CEI sets (FA-27, SA-22, and SM-16 CEMs).

In the experiments described below, we used the PubMed database [15]. The PubMed search was conducted using Entrez [42], an integrated database retrieval system that allows access to a diverse set of databases at the National Center for Biotechnology Information (NCBI) [43] website. The published articles that were obtained through search of PubMed are read using the REACH engine [19], which extracted a list of events and the corresponding information (see Section 2.1). The REACH reading engine is available online and can be run through the Integrated Network and Dynamical Reasoning Assembler (INDRA) [44]. We conducted our analysis on three different CEI sets, which were obtained using varying levels of automation and manual intervention. For each set, the list of events, with associated entities, is automatically translated from the reading output into BioRECIPES tabular format.

In the *fully automated* (*FA*) approach, both the PubMed database search for relevant articles and the extraction of event data from the selected articles were done by machines. Specifically, in the FA experiment, we used search query "T-cell and (PTEN or AKT or FOXO)" and selected top 11 from the best matched papers, by the PubMed search engine. In the *semi-automated* (*SA*) approach, we selected papers that are cited by [13] and used the event information that REACH extracted from those papers. Finally, in the *semi-manual* approach (*SM*), we rely the most on human intervention, we manually excluded from the SA reading output those interactions that violate any assumptions made by the authors originally in [12]. For instance, the authors in [12] consider element TCR to be an input to the network, and therefore, TCR should not have any regulators in the T cell model. Therefore, if REACH retrieves an interaction in which TCR is a regulated element, we manually remove these interactions and keep only the interactions having TCR as a regulator.

Model extension algorithms are written in Python. The statistical model checker is written in C++ and it was used to test all CEMs on a set of properties listed in the supplementary material (Figure S4). The properties are written as BLTL formulas. The overall iterative model extension tool is written in Python, and it was run on a 3.3 GHz Intel Core i5 processor. For the clustering algorithm, we used the MCL package from [32], and for the visualization of the graphs we used Cytoscape [35].

## 5.2 Role of static network characteristics

In **Figure 2**(a), we show three networks (undirected interaction maps, for easier visualization) that were formed by combining each of the CEI sets (FA, SM, and SA) with the baseline model. In other words, as described in Section 3.2, we created graphs $G^{new,FA}(V^{new,FA}, E^{new,FA})$, $G^{new,SA}(V^{new,SA}, E^{new,SA})$, and $G^{new,SM}(V^{new,SM}, E^{new,SM})$. We explored static characteristics of these three graphs and their correlation with the selection of the best extended model. Given a directed graph $G^{new,*}(V^{new,*}, E^{new,*})$, where $* \in \{FA, SA, SM\}$ and the definition of a path in Section 3.2, we computed average path length (*APL*), clustering coefficient (*Coeff*), and graph density (*D*) [45]. Assuming that a distance $d(v_i, v_j)$ is the number of edges on a shortest path between nodes $v_i$ and $v_j$, *APL* is computed as an average distance across all possible pairs of nodes in the graph:

$$APL = \frac{1}{|V^{new,*}| \cdot (|V^{new,*}| - 1)} \cdot \sum_{v_i, v_j \in V^{new,*}, v_i \neq v_j} d(v_i, v_j)$$

where $|V^{new,*}|$ is the number of nodes in the graph. If there is no path between $v_i$ and $v_j$, then $d(v_i, v_j) = 0$. The *clustering coefficient* (*Coeff*) [45] is computed for each node $v_i$ in a directed graph as:

$$Coeff(v_i) = \frac{T(v_i)}{degree^{tot}(v_i) \cdot (degree^{tot}(v_i) - 1) - 2 \cdot degree^{\leftrightarrow}(v_i)}$$

where $T(v_i)$ is the number of triangles (three connected nodes) in the graph that contain node $v_i$, $degree^{tot}(v_i)$ is the sum of the in-degree (the number of incoming edges) and the out-degree (the number of outgoing edges) of $v_i$, and $degree^{\leftrightarrow}(v_i)$ is the reciprocal of $degree^{tot}(v_i)$. *Coeff* is a number between 0 and 1, and therefore, if an average *Coeff* value, computed across all graph nodes, approaches 0, the graph is more likely to contain stars, while when this value approaches 1, the graph is a clique. The *graph density* (*D*) [45] is defined for a directed graph as:

$$D = \frac{|E^{new,*}|}{|V^{new,*}|(|V^{new,*}|-1)}$$

where $|E^{new,*}|$ is the number of edges and $|V^{new,*}|$ is the number of nodes in the graph. A graph is considered to be dense if the number of edges is close to the maximum number of possible edges, therefore, the graph density is close to 1 for a dense graph and close to 0 for a sparse graph.

We list in the table in **Figure 2**(a) the main parameters for graphs $G^{new,FA}$, $G^{new,SA}$, and $G^{new,SM}$. As can be seen from the table, the $G^{new,FA}$ has the largest number of nodes and edges, and it results in the largest number of clusters. On the other hand, $G^{new,SA}$ and $G^{new,SM}$ have smaller number of edges and nodes. In **Figure 2**(b), we highlight the difference between the three graphs: $G^{new,FA}$ is shown in green, $G^{new,SA}$ in blue, and $G^{new,SM}$, which is a subgraph of $G^{new,SA}$, in orange. In addition, we show the overlapping nodes between the three networks in cyan.

Interestingly, it was observed that despite network diversity, $G^{new,FA}$, $G^{new,SA}$, and $G^{new,SM}$ share prominent structural features: they have small *APL*, small average *Coeff*, and small *D*, and thus, large average *degree^{tot}* values are unlikely. This similarity is even better illustrated in **Figure 2**(c), showing the *degree^{tot}* histogram for the nodes in each network that follows a power law, and the distribution of node distance (*d*) centered approximately around value 4. As can be noticed, both graph parameters, *degree^{tot}* and *d*, have similar patterns but with different count numbers for each $G^{new,*}$ in proportion to the size of its network. Moreover, the values of *D* in the table in Figure 2(a) suggest that the graphs assembled from the information extracted by machine readers are less dense, even with varying network size. These results also suggest that the difference in literature sources and the size of the CEI sets did not affect the characteristics of $G^{new,*}$ graphs in our case study.

Following our Algorithm 1, shown in Figure S1 in the supplementary material, we clustered the three graphs $G^{new,*}$. We obtained 22 clusters from $G^{new,FA}$ ($C_1^{FA}, C_2^{FA}, \ldots, C_{22}^{FA}$), 11 clusters from $G^{new,SA}$ ($C_1^{SA}, C_2^{SA}, \ldots, C_{11}^{SA}$), and 9 clusters from $G^{new,SM}$ ($C_1^{SM}, C_2^{SM}, \ldots, C_9^{SM}$). The inspection of obtained clusters shows that they are less dense and star-like networks (two examples shown in **Figure 2**(d)), which agrees with the conclusions of the above studies of graph characteristics. Thus, computing the graph parameters can guide our proposed extension method by providing an early estimate of whether the CEI sets can lead to desired models. For instance, if the *APL* is large, we will expect to extract a fewer number of return paths (defined in Section 3.2) from the $G^{new,*}$ graphs, and therefore, in our analysis we will lack the connectivity of the CEIs to the baseline model. Additionally, the graphs with smaller *D* will reduce the computation time, and computing this parameter helps determine in advance the expected execution time of our algorithm.

### 5.3 Assembly and evaluation of T cell CEMs

To extend the baseline model, we first test the connectivity of each cluster to the model by searching for a return path (starts and ends in the baseline model) between an individual cluster and the model. In **Figure 2**(d), we highlight in blue a return path that exists between cluster $C_1^{SM}$ and the baseline model (TCR → AKT → MTORC2), and a return path that exists between cluster $C_2^{SM}$ and the baseline model (TCR → NEDD4_ext → PTEN), where clusters $C_1^{SM}$ and $C_2^{SM}$ are two of the nine clusters generated from $G^{new,SM}$. Furthermore, we explored multiple cluster connectivity with respect to return paths and created 5, 11, and 7 additional CEMs by merging two clusters together from the clusters obtained from $G^{new,FA}$, $G^{new,SA}$, and $G^{new,SM}$ graphs respectively. Therefore, the total number of CEMs resulting from the FM, SA, and SM CEI sets are 27, 22, and 16, respectively. We also highlight in red in **Figure 2**(d) a return path that exists between the baseline model, and clusters $C_1^{SM}$ and $C_2^{SM}$ (PI3K → PIP3 → AKT → Foxo1_ext → PTEN).

The final list of CEMs includes the baseline model extended by one or two of the clusters generated by MCL for each $G^{new,*}$ graph. For several candidate models, and for all the 27 properties, we show in **Figure 2**(e) the global CEM probability (defined in Section 3.4) for all CEMs, in the FA, SA, and SM cases. The highest peak per graph represents the best candidate model when compared to the golden model. From the morphology of the generated clusters, we expect a cluster to affect the behavior of the model if it contains the key elements included in system properties. As a consequence, the estimated probability (*P*) values will vary for those clusters. However, clusters lacking those key elements will most probably not affect the behavior of the model, and thus, larger number of properties will not be satisfied. In our case study, we found that the CEMs that include two clusters with key elements satisfy a larger number of properties. Merging clusters helped increase the probabilities, however, the 70% steady-state level of Foxp3 in the low-dose scenario observed in [13] is not achieved. The best performance is obtained for the model that combines two clusters, $C_1^{SM}$ and $C_2^{SM}$ (Figure 2(d)), which satisfies almost all properties (24 out of 27), (Figure 2(e)). Additionally, these two clusters together restored all the missing interactions removed from the golden model (Section 4.1). The network of the best CEMs for the SM CEI set is shown in Figure S5 in the supplementary material.

### 5.4 Guided extension of executable models

A model created automatically with ACCORDION, using the information from the papers in public databases, which satisfies most of the desired properties, may not be the same as the golden model. The differences can be found in both network structure and element update rules. With our extension methodology, we sometimes obtain multiple models that satisfy the same set of properties. This diversity helps us examine redundancies or discover alternative pathways regulating the same target element.

Adding new elements to the baseline model requires the consideration of three factors, cluster granularity, regulatory functions, and initial values. First, when using MCL to cluster our directed networks ($G^{new,*}$), the principal handle for changing cluster granularity is the inflation parameter, *r*, described in Section 3.2. An increase in *r* causes an increase in the cluster granularity. Therefore, we explored the effect of *r* on finding the best set of clusters for each CEI set. In [32], the authors determined a good interval to choose from (e.g., from 1.1 to 10.0), however, the range of suitable values will certainly depend on the input graph. For our case study and the different reading output sets, we found that *r* = 1.1 is too low, and *r* ≥ 6.0 is too high. We have therefore chosen value *r* = 4 for our studies and conducted experiments based on this value (**Figure 3**(a)).

Additionally, MCL is usually applied to undirected graphs, and thus, if there is an edge between nodes $v_i$ and $v_j$, the corresponding entries in the adjacency matrix will be $s(i,j) = s(j,i) \neq 0$. Since we use directed graphs, an edge from node $v_i$ to node $v_j$ would correspond to the $s(i,j)$ entry, but not to the $s(j,i)$ entry. To solve this issue, we make $s(j,i)$ equal to $s(i,j)$, which seems to ignore an important information (directionality). However, the directionality does not affect the static structure of clusters [32]. Finally, the runtime of the extension algorithm is proportional to the number of properties that we need to test against. In other words, if we have $N_C$ clusters and $N_P$ properties, the time required for the extension algorithm is at the order of $O(N_C*N_P)$. However, the time complexity can be reduced to $O(N_C)$ if testing for all the clusters is carried out in parallel.

The second challenge is in deciding which operations to use (e.g., AND or OR for Boolean functions), when adding new element regulators found in CEIs. We investigated element update rules, and the difference between our final model and the golden model. For instance, the PTEN update rule from the model in [13] is PTEN = (FOXO1* MEK1) * (!CK2 + !NEDD4). As our baseline model starts with PTEN = (FOXO1 * MEK1), we need to "recover" the rest of the update rule from the CEI set, that is, negative regulation of PTEN by CK2 and negative regulation of PTEN by NEDD4. In the process of adding new element regulators, if the information about the regulation type or the importance of the regulator (e.g., necessary vs. sufficient) is available, it will guide the operation choice (AND vs. OR).

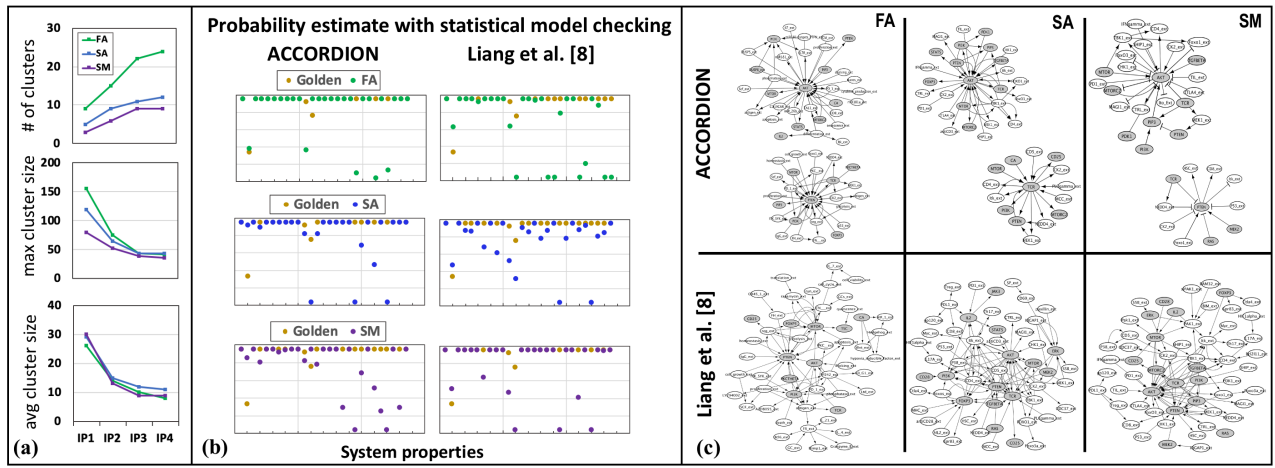

**Figure 3.** (a) Several cluster characteristics measured as functions of inflation parameter (IP), for the FA, SA, and SM cases (IP1=0.5, IP2=2, IP3=4, IP4=6). (b) Comparison of the probability estimate $P$ for all 27 properties, for the golden model, and for the best model obtained from each of the three CEI sets (FA, SA, SM) using ACCORDION and the method from [8]. (c) Clusters that were included in the final CEMs for the FA, SA, and SM cases, for ACCORDION and [8] (white nodes are new nodes from the CEIs and gray nodes are the nodes in the CEIs that also exist in the baseline model).

The third challenge is in deciding elements' initial values (e.g., 0 or 1 in Boolean models) when simulating CEMs and testing their behavior against the desired system properties. We found that assigning different initial values to those source and target elements of CEIs that were not in the original baseline model have quite similar results. This emphasizes the robustness of the baseline model and that the final extended model is influenced by the values of elements in the baseline model. These findings are mostly in agreement with what has been shown in [8]. It is likely that these results are influenced by our choice of the case study, and the fact that we defined system properties (Figure S4 in the supplementary material) in terms of steady states of the key model elements. As our next steps, we will further explore effects of initialization, as well as add system properties with more complicated temporal relationships between elements.

### 5.5 Comparison with previously proposed methods

We tested the effectiveness of the previously proposed model extension method from [8] when applied to our case study. This is achieved by replacing our model extension method with the method introduced in [8], using the same baseline T cell model (described in Section 4.1) and the three CEI sets (FA, SA, and SM). In [8], the authors described an automated extension method that considers only the extensions that are connected to the baseline model. They first identify a set of baseline model elements of interest, and then add the extensions based on the proximity to these elements. The proximity is measured as a number of edges on a path connecting baseline model elements and new elements in extensions, and the extension is conducted in layers, starting from the baseline model. Several extension configurations are proposed in [8], depending on the extension approach that the user could be interested in. For example, the focus of model extension can be including the regulation of a certain element or a set of elements, regardless of the number of extension layers this would require. Another approach discussed in [8] focuses on reducing the number of layers while tracking the effect of adding new extensions to the baseline model. In this work, we focus on studying the effect of adding new extensions to the baseline model, therefore, we used the latter approach from [8] in the comparisons.

Figure 3(b) highlights the differences between the results of our method and the method from [8], when tested using statistical model checking. We compared the $P$ values for each property and each CEM for the two methods. As can be observed, ACCORDION outperforms the method from [8] in the case of the FA and SA CEI sets. However, in the SM case, the method from [8] shows slightly better results. These results indicate that the layer-based approach is less effective when used on a large set of CEIs and without any human intervention. The visualization of the topology of the sets of extensions extracted by each method is shown in Figure 3(c). Our method provides concise groups of connected CEIs, that are at the same time connected to the baseline model through return paths. On the other hand, the networks generated by the method from [8], show several nodes that are extensions and are downstream from the baseline model, which means they are not affecting the baseline model (**Figure 3**(c)). Thus, the comparisons we conducted suggest that the Liang et al. method [8] has two major limitations that our method overcomes: it is subjective and prone to human judgment variation in selecting the number of elements of interest and the number of layers, and it becomes impractical with the large number of layers.

## 6 Conclusion and future work

In this paper, we have described a novel methodology and a tool, ACCORDION, that can be used to automatically assemble the information extracted from literature into models. Our proposed approach combines machine reading with clustering, simulation, and model checking, into an automated framework for rapid model assembly and testing to address biological questions. Furthermore, by automatically extending models with the information published in literature, our methodology allows for efficient collection of the existing information in a consistent and comprehensive way, while also facilitating information reuse and data reproducibility, and replacing hundreds or thousands of manual experiments, thereby reducing the time needed for the advancement of knowledge. As our future work, we will explore opportunities to improve ACCORDION. For example, we plan to utilize the information about corroborations and contradictions to inform clustering and CEM selection, and we will work on parallelization of our algorithms to further increase their execution efficiency.

## Funding

This work is supported by DARPA grants W911NF-17-1-0135 and W911NF-18-1-0017 awarded to N. Miskov-Zivanov.